\documentclass[reprint,longbibliography,
 amsmath,amssymb,superscriptaddress]{revtex4-2}
\usepackage{graphicx}
\usepackage{dcolumn}
\usepackage{bm}
\usepackage[colorlinks=true, citecolor=blue, linkcolor=blue, urlcolor=blue]{hyperref}
\usepackage{orcidlink}
\usepackage{xcolor}
\usepackage{amsmath, amssymb, amsfonts}
\usepackage{mathtools}
\usepackage{subfigure}
\usepackage[normalem]{ulem}
\usepackage{mathrsfs}
\usepackage{sidecap}
\usepackage{verbatim}

\definecolor{purple1}{rgb}{128,0,128}

\usepackage{bigints}
\newcommand{\bea}{\begin{eqnarray}}
\newcommand{\ea}{\end{eqnarray}}

\definecolor{darkpastelgreen}{rgb}{0.01, 0.75, 0.24}

\def\x{\mathbf{x}}

\def\k{\mathbf{k}}

\def\d{\mathrm{d}}

\begin{document}

\title{Effective spacetime description of light propagation in linear magnetoelectric media}

\author{Lucas T. \surname{de Paula}\orcidlink{0009-0002-1192-3532}}
\email{tobias.l@ufabc.edu.br}
\affiliation{Centro de Matemática, Computação e Cognição, Universidade Federal do ABC, \\
Santo Andr\'e, S\~ao Paulo 09210-170, Brazil}

\author{Caio C. \surname{Holanda Ribeiro}\orcidlink{0000-0001-6571-4168}}
\email{caiocesarribeiro@alumni.usp.br}
\affiliation{International Center of Physics, Institute of Physics, University of Brasilia, 70297-400 Brasilia, Federal District, Brazil} 

\author{Vitorio A. \surname{De Lorenci}\orcidlink{0000-0001-5880-2207}}
\email{delorenci@unifei.edu.br}
\affiliation{Instituto de F\'{\i}sica e Qu\'{\i}mica, Universidade Federal de Itajub\'a, \\
Itajub\'a, Minas Gerais 37500-903, Brazil}

\begin{abstract}

Formal analogies between gravitational and optical phenomena have been explored for over a century, providing valuable insights into kinematic aspects of general relativity. Here, this analogy is employed to study light propagation in linear magnetoelectric media from an effective spacetime perspective. Starting from Maxwell’s equations in covariant form, it is shown that an effective metric can always be identified for linear, non-dispersive magnetoelectric materials. The effective metric is then used to construct analog models in the limit of geometric optics. Among the optical effects analyzed, it is shown that under reasonable assumptions on the magnitude of the magnetoelectric response, a one-way propagation region can be established, which behaves analogously to an event horizon.


\end{abstract}

\maketitle

\setlength{\parskip}{0pt}
\section{Introduction}
\label{one}
Analog models for curved space phenomena have been a subject of investigation since the beginning of the 20th century, when Gordon originally studied light propagation in material media and reinterpreted the refractive index of a medium by means of an effective geometric description \cite{gordon}. Throughout the years, the possibility of creating analogs of spacetime geometries in laboratories was extensively studied \cite{barcelo3, Barcelo2019}, not only in the realm of electromagnetism \cite{tamm,plebanski,plebanski2,novello,PhysRevLett.128.091301,PhysRevA.89.053807,PhysRevA.104.043523,PhysRevD.110.024035,PhysRevA.93.053820,Ribeiro2020}, but also in the context of acoustic waves and condensed matter systems \cite{unruh, barcelo2,fabbri,PhysRevB.86.144505,Unruh1981}. Models containing an event horizon in Bose-Einstein condensates have also been frequently examined \cite{barcelo,fabbri2,PhysRevA.76.033616,PhysRevD.107.L121502,Jacobson2018,Garay,Jeff2019,tedbhlaser}, which includes the analysis of analog Hawking's radiation phenomenon. 
In addition, analog models seem to be interesting tools to test metric solutions related to controversial predictions, such as those containing closed time-like curves \cite{Vilenkin:2000jqa, lakhtakia,barceloctc} (even though quantum physics suggests that such possibilities are forbidden \cite{hawkingctc,edsom}). These studies highlight the versatility of analog models and motivate the exploration of new physical systems, particularly in the realm of advanced optical materials.

Recent advances in the science and technology of optical materials, which include magnetoelectrics \cite{fiebig, rivera, schmid, Tabares,Zhai2017,Oh2014,Fujii2022} and metamaterials \cite{lapine2014}, have opened a new window for investigating analog models using electromagnetic phenomena.
In particular, in a magnetoelectric material, the polarization phenomenon can be induced by a magnetic field, and magnetization can be induced by an electric field, or both together.
%

%
%

In this work, light propagation in linear magnetoelectric media within an effective spacetime framework is investigated. Phenomena associated with specific spacetime configurations are identified and discussed. By assuming a regime where dispersion can be neglected, i.e., the delay of the material response to an external electromagnetic field is negligible, it is shown in a fully covariant manner that a general linear magnetoelectric material leads naturally to the notion of a rank-two symmetric tensor that can be interpreted as an effective metric tensor pertaining to a certain curved space. Furthermore, it is shown how this effective metric governs light ray propagation in the regime of geometric optics.
As a consequence of the geometric optics analysis, a family of analog models containing surfaces that can be crossed by light in only  one direction is built. It is shown that for a naturally occurring magnetoelectric material, an analog description can be constructed in which light rays experience two distinct propagation barriers: one acting as a true one-way surface, analogous to an effective event horizon, and another that partially restricts propagation depending on the polarization mode. Moreover, by performing an appropriate coordinate transformation, it is shown that the linear magnetoelectric system can be mapped to a locally flat spacetime, enabling the exploration of analog models inspired by general relativity. In particular, this mapping reveals a formal connection with the spinning cosmic string, a solution that admits closed time-like curves \cite{Vilenkin:2000jqa}. We assess this time-travel–like behavior and its relation to the cosmic string through a toy model in which magnetoelectric effects dominate over the ordinary refractive index of the material; within this framework, however, we discuss that such a time-travel mechanism cannot be physically assessed by means of the optical material.

The work is organized as follows. In next section, Maxwell's equations are written in a covariant form for a linear dispersionless magnetoelectric material in terms of a symmetric rank-two tensor built from the optical coefficients. 
Section \ref{two} determines  how the effective metric interpretation occurs, and the eikonal equation in terms of the effective metric is deduced within the regime of geometric optics. In Sec.~\ref{analog} concrete examples of spacetime analogs are presented, and the solutions to the eikonal equation are illustrated.
Final remarks are presented in Sec.~\ref{discussion}. Throughout the text, we adopt the abstract index notation and the metric signature convention of \cite{wald2010general}, for which the first Latin letters, $a,b,c,\ldots$, are merely an indication of the tensor-type and rank. For instance, $E^{a}$ is a contravariant tensor of rank $1$. Greek letters $\alpha , \beta,\gamma,\ldots$, run from 0 to 3 and denote tensor components in a given base, whereas Latin letters from the middle of the alphabet, $i, j, k,\ldots$, run from 1 to 3, and denote the components of $3$-vectors. Einstein's convention for repeated indices is used, and units are such that $G=c=\epsilon_0=1$.

\section{Maxwell's equations in the presence of magnetoelectric media}
\label{secmax}

Electromagnetic phenomena is in general described by the field $A_a$ by means of the field strength tensor $F_{ab}=\partial_aA_b-\partial_bA_a$. In vacuum and in the absence of sources, Maxwell's equation are simply
\begin{equation}
    \partial_aF^{ab}=0,
\end{equation}
where the Minkowski metric $\eta_{ab}$ is used to raise and lower tensor indices. In the presence of polarizable and magnetizable matter, however, the above equation assumes a different form, which we discuss now. 

Let $v^{a}$, with $v^av_a=-1$, be the velocity field of a family of observers that are locally at rest with respect to the matter distribution. Accordingly, because $v^a$ is time-like, it defines a notion of spatial vectors (and tensors) through $A^a=(h^a_{\ b}-v^av_{b})A^b$ via the projection operator $h^{a}_{\ b}=\delta^{a}_{\ b}+v^av_b$. Thus the electric field $E_a$ and the magnetic field $B^{a}$ locally experienced by matter are \cite{wald2010general}
\begin{align}
    E_a&=F_{ab}v^{b},\\
    B^{a}&=-\frac{1}{2}\epsilon^{abcd}F_{bc}v_d,
\end{align}
where $\epsilon_{abcd}$ is the Levi-Civita tensor defined (in a local coordinate system) by  $\epsilon_{0123}=\sqrt{-\eta}$, $\eta$ being the determinant of the metric tensor. Note that both $E_{a}$ and $B^a$ are spatial vectors. In addition to enable the correct identification of the fields, it is straightforward to show that
\begin{equation}
    F^{ab}=2v^{[a}E^{b]}-\epsilon^{abcd}B_cv_d.
\end{equation}

When polarization and magnetization occur, the (averaged) Maxwell's equations assume the form
\begin{equation}
    \partial_aP^{ab}=0,
\end{equation}
where the antisymmetric tensor $P^{ab}$ is defined in terms of the auxiliaxy fields $D^{a}, H_{a}$ as
\begin{equation}
    P^{ab}=2v^{[a}D^{b]}-\epsilon^{abcd}H_cv_d,\label{eqP}
\end{equation}
together with constitutive relations connecting $D^{a}$ and $B^{a}$ to $E^a$ and $H^a$.

For definiteness, we assume a non-dispersive, linear magnetoelectric material, described by the constitutive equations
\begin{align}
    D^{a}&=\varepsilon h^{ab}E_b+\epsilon^{abcd}H_{b}v_c\alpha_d,\label{const1}\\
    B_a&=\mu h_{ab}H^{b}-\epsilon_{abcd}E^{b}v^c\alpha^d,\label{const2}
\end{align}
where $\alpha^a$ is an arbitrary spatial vector (with respect to $v^a$). It can be shown that, using the above constitutive relations, Eq.~\eqref{eqP} can be rewritten as
\begin{equation}
    P^{ab}=\frac{n}{\mu}g^{ac}g^{bd}F_{cd},\label{auxfinalremarks}
\end{equation}
where we have defined the tensor
\begin{equation}
    g^{ab}=\frac{1}{\sqrt{n}}\left[\eta^{ab}-(n^2-1-\alpha^2)v^av^b+v^a\alpha^b+v^b\alpha^a\right].\label{effectivemetric}
\end{equation}
Here, $\alpha^2=\alpha_a\alpha^a$ and $n=\sqrt{\varepsilon\mu}$ is the medium refractive index. Finally, the dynamics of the electromagnetic field is governed by the equation
\begin{equation}
    \partial_a\left(\frac{n}{\mu}g^{ac}g^{bd}F_{cd}\right)=0,
    \label{maxfinal}
\end{equation}
together with the definition of the tensor $F_{ab}$ in terms of the vector potential. 

In order to obtain a deeper insight about the constitutive relations Eq.~\eqref{const1} and \eqref{const2}, let us show that they produce the expected magnetoelectric effect for matter at rest. Let $\{t,x,y,z\}$ be the standard Cartesian coordinates. Matter at rest with respect to the laboratory has as velocity field $v^{a}=(\partial_t)^{a}$, which becomes $v^{\mu}=(1,0,0,0)$. In this case, we find that $E_0=B_0=D_0=H_0\equiv 0$ and $\alpha_0=0$, and thus only spatial components are relevant in Eqs.~\eqref{const1} and \eqref{const2}. For instance, $E_{\mu}=(0,E_1,E_2,E_3)$, where $E_i$, $i=1,2,3$, are the Cartesian components of the electric field, and similarly for the other fields. Therefore, a straightforward computation shows that the constitutive relations assume the form
\begin{align}
    D_i&=\varepsilon E_i+\sum_{j,k}\epsilon_{ijk}H_j \alpha_k,\\
    B_i&=\mu  H_i-\sum_{j,k}\epsilon_{ijk}E_j \alpha_k,
\end{align}
where $\epsilon_{ijk}=\epsilon_{0ijk}$. The equations above reproduce the magnetoelectric effect presented in \cite{landau} when the antisymmetric tensor $\alpha_{ij}=\sum_{k}\epsilon_{ijk}\alpha_k$ is identified. Naturally, solving Maxwell's equations in the form \eqref{maxfinal} or in standard Cartesian form produces the same results for matter at rest, and the advantage of the form \eqref{maxfinal}, in addition to its covariance, is the identification of the tensor $g^{ab}$ in terms of an arbitrary velocity field, as we discuss in the next section.  

\section{Analog spacetime}
\label{two}

Equation \eqref{maxfinal} is supposed to describe all electromagnetic phenomena in non-dispersive, linear magnetoelectric materials in arbitrary movement and in the absence of spacetime curvature. Nevertheless, the notation chosen for the tensor $g^{ab}$ already suggests a connection to gravity, which we discuss now. 

Consider a local coordinate system \(x^{\mu}\). We define a new tensor field \(g_{\mu\nu}\) as the inverse of \(g^{\mu\nu}\), such that
\begin{equation}
 g^{\mu\alpha} g_{\alpha\nu}  = \delta_\nu^\mu.
\label{1}
\end{equation}
At this point, it is worth emphasizing that the background spacetime geometry is here assumed to be the Minkowski metric, $\eta_{\mu\nu}$. Thus, there are two distinct notions of covariant tensor associated with $ g^{\mu\nu}$, namely, the tensor $g_{\mu\nu}$ defined by Eq.~\eqref{1},
\begin{align}
    g_{\mu\nu} = \sqrt{n}\left[\eta_{\mu\nu} + v_\mu v_\nu -\frac{1}{n^2}(v_\mu + \alpha_\mu) (v_\nu + \alpha_\nu) \right],
    \label{metric_dow}
\end{align}
and the one defined by $\eta_{\mu\alpha}\eta_{\nu\beta}  g^{\alpha\beta}$. They coincide only when $\epsilon=\mu=1$, and $\alpha^{\mu}=0$.

 We note the property
\begin{equation}
    \det(g^{\mu\nu})=\det(\eta^{\mu\nu}),
\end{equation}
such that
\begin{equation}
    g\equiv\det(g_{\mu\nu})=\eta,
\end{equation}
from which Eq.~\eqref{maxfinal} can be cast in its  equivalent form
\begin{equation}
    \nabla_\alpha\left(\frac{n}{\mu}g^{\alpha\rho}g^{\beta\sigma}F_{\rho\sigma}\right)=0.\label{maxfinal2}
\end{equation}
The covariant derivative operator in the above comes from the Levi-Civita connection defined with respect to the ``effective metric'' $g_{\mu\nu}$. We note that when $\nabla_\alpha(n/\mu)\equiv 0$, e.g., for homogeneous materials, the field equation \eqref{maxfinal2} is identical to the field equation for the minimally coupled electromagnetic field in a curved space described by the metric $g_{\mu\nu}$, whereas a non-vanishing $\nabla_\alpha(n/\mu)$ models non-minimal coupling to gravity in this fictitious spacetime \cite{Ribeiro2020}. It is worth mentioning that the possibility of expressing the auxiliary tensor $P_{\mu\nu}$ in terms of the effective optic metric $g_{\mu\nu}$, such that the field equations take the form shown above, has already been explored in the case of non-magnetoelectric anisotropic media \cite{Ribeiro2020}. Moreover, a similar approach based on nonlinear electrodynamics has recently been studied \cite{Goulart:2025tjn}. In general, it follows from the results of \cite{Ribeiro2020} that Eq.~\eqref{maxfinal2} also holds for nonlinear theories for which nonlinear effects can be modeled by field-dependent susceptibilities.

Although Eq.~\eqref{maxfinal2} describes all possible phenomena 
 (excepting stress tensor related effects) regarding electromagnetic disturbances in the system, we shall focus on its solutions in the limit of geometric optics. Specifically, we are interested in solutions whose spacetime variation occurs rapidly in comparison to any other scale. We can write such solutions in the form \cite{wald2022}
 \begin{equation}
     A_{\mu}=\left[\sum_{n=0}^\infty a_{\mu}^{(n)}\epsilon^n\right]e^{iS/\epsilon},\label{ansatz}
 \end{equation}
where $S=S(x^\mu)$ is the phase, and $\epsilon>0$ is a small dimensionless constant added to distinguish the different scales in the system. This ansatz can be substituted in either Eq.~\eqref{maxfinal} or \eqref{maxfinal2}, and the results are clearly the same. By working in the Lorenz gauge, $g^{\mu\nu}\nabla_\mu A_\nu=0$, we find that at leading order in powers of $\epsilon$ that $g^{\mu\nu}k_\mu a^{(0)}_{\nu}=0$, whereas Eq.~\eqref{maxfinal2} implies the eikonal equation
\begin{equation}
    g^{\mu\nu}k_\mu k_\nu=0,\label{nullcondition}
\end{equation}
and $k_\mu=\partial_\mu S$ is the wave vector. Higher order terms give rise to the equations satisfied by the amplitudes $a^{(n)}_\mu$. For instance, the next order equations for both the Lorenz gauge condition and the field equations are $g^{\mu\nu}\nabla_\mu a^{(0)}_\nu=-ik^\mu a^{(1)}_{\mu}$ and
\begin{align}
    2k^\alpha\nabla_\alpha& a^{(0)}_\nu+a^{(0)}_\nu \nabla_\alpha k^\alpha\nonumber\\
    &+\left(\nabla_\alpha \ln\frac{n}{\mu}\right)g^{\alpha\rho}\left[k_\rho a^{(0)}_\nu-k_\nu a^{(0)}_\rho\right]=0,\label{auxamplitude}
\end{align}
where we defined $k^{\mu}=g^{\mu\nu}k_{\nu}$. Note that $k^{\mu}$ is in general different from the contravariant wave vector obtained from $\eta^{\mu\nu}$. Contracting the above equation with $a^{(0)}_{\rho}g^{\rho\nu}$ produces the continuity equation
\begin{equation}
    \nabla_\alpha\left[\frac{n}{\mu}g^{\rho\nu}a^{(0)}_\rho a^{(0)}_\nu k^{\alpha}\right]=0,
\end{equation}
which furnishes the particle-like interpretation for light rays known from geometric optics \cite{wald2022}. Therefore, solutions in form \eqref{ansatz} are fully characterized by the eikonal equation \eqref{nullcondition}, $k^{\nu} k_\nu=0$, that is, $k_{\mu}$ is a null vector in the effective spacetime. Note that, in the case where $\varepsilon=\mu=1$, $\alpha^\mu=0$ (vacuum), the above equation is simply $\eta^{\mu\nu} k_\mu k_\nu  =0$. We note also an important constraint imposed by geometric optics. By assumption, the amplitude of the solutions is slow varying with respect to their phases. This should be kept in mind in Eq.~\eqref{auxamplitude}, that is a differential equation for the dominant term of the field amplitude and contains derivatives of the optical coefficients. In the case where the material properties vary considerably in comparison to the involved wavelengths, the geometric optics approximation might not be reliable.
%
%

Therefore, in an optical material the dispersion relation assumes a more elaborated structure in comparison to its empty space counterpart. The optical properties of the medium are added to the analysis in such a way that the dispersion relations generalizes to $(\eta^{\mu\nu} + \theta^{\mu\nu} )k_\mu k_\nu=0$, where $\theta^{\mu\nu}$ is related to the susceptibilities coefficients of the medium via Eq.~\eqref{effectivemetric}. As a consequence, the magnitude of the phase velocity of light in a material medium will be generally dependent on its optical susceptibilities, the applied fields and also on the direction of wave propagation. 
%
%
%
%

For our purposes, a key aspect of the null condition \eqref{nullcondition} is that it implies that $k_\mu$ satisfies the geodesic equation for light rays in the effective spacetime \cite{ndke2001,prdklippert2002, previtorio2002}.  Indeed, it follows  from the notion of covariant derivative linked to $g_{\mu\nu}$ and Eq.~\eqref{nullcondition} that 
\begin{align}
g^{\mu\nu}k_{\mu}\nabla_\alpha k_{\nu}=0,
\end{align}
and because $\nabla_\alpha k_{\nu}=\nabla_\alpha \nabla_{\nu}S=\nabla_\nu \nabla_{\alpha}S=\nabla_\nu k_\alpha$, the above equation is equivalent to
\begin{equation}
    k^{\mu}\nabla_\mu k^{\alpha}=0.\label{geodesic}
\end{equation}
The equation above assumes its customary form when written in terms of the integral curves of $k^{\mu}$, defined by the condition
\begin{equation}
    \frac{\d x^{\mu}}{\d \lambda}=k^{\mu},
\end{equation}
where $x^{\mu}=x^{\mu}(\lambda)$ and $\lambda$ is an affine parameter. In particular, Eq.~\eqref{nullcondition} implies
\begin{equation}
    0=k^{\mu}k_{\mu}=\frac{\d x^{\mu}}{\d \lambda}\frac{\partial S}{\partial x^{\mu}}=\frac{\d S}{\d\lambda},\label{phasecons}
\end{equation}
that is, the integral curves of $k^{\mu}$ are curves of constant phase $S$ and thus can be used to determine how wave fronts evolve (with respect to the laboratory time) in the analog optical system.

%
%
%
%
%

Notice that Eq.~\eqref{geodesic} is the geodesic equation in the spacetime described by the geometry ${g}_{\mu\nu}$. 
Thus, in the limit of {\it geometric optics and in the absence of dispersion}, a light ray propagating in a material medium can be used to model light rays propagating in a curved spacetime, which is a solution of general relativity. This is a mathematical equivalence that holds as far as kinematic aspects of general relativity are considered. Also, it is important to note that the geometric optics approximation breaks down in the vicinity of caustics, where the amplitude of the fields are expected to vary rapidly.

%
%

It is worth emphasizing that all the above results did not make use of $g_{\mu\nu}$ as the spacetime metric. It is just an effective one that is experienced only by the wave vector. The spacetime under consideration is the Minkowski space.

\section{Light ray propagation and analog spacetime effects}
\label{analog}

Let us now study light propagation in particular setups and investigate the possibility of relating the optical phenomena to specific kinematic features of curved spacetimes. We restrict our analysis to the case where the material is at rest with respect to the laboratory frame, that is, if $t$ is the laboratory coordinate time, we let $v^{a}=(\partial_t)^{a}$.

Let us consider the case of a slab-like magnetoelectric material. By working with the Cartesian coordinates $\{t,x,y,z\}$, we assume
\begin{equation}
\alpha^{\mu}=(0,\alpha,0,0),   
\end{equation}
where $\alpha$ and the refractive index $n$ are functions of $x$ alone, and $\alpha$ is non-vanishing only around $x=0$, the position of the slab.

We consider first a realistic scenario from an experimental perspective, namely, we assume $n\sim 1$ and $|\alpha|\ll1$, which is expected to occur in most magnetoelectric materials. Instead of working with the geodesic equations, let us treat the eikonal equation, \eqref{nullcondition}, directly. We look for solutions for the eikonal in the form $S=-\omega t+F(x,y)$,
that corresponds to light rays propagating in the $xy$ plane. Equation \eqref{nullcondition} then becomes
\begin{equation}
    (k_x-\alpha\omega)^2+k_y^2=n^2\omega^2,\label{tosolveBH}
\end{equation}
and $\k=\nabla F$ is the vector orthogonal to the surfaces of constant $S$ at a fixed time. Due to the spatial translation symmetry in the $y$ direction, solutions to the eikonal might be taken with $F=k_y y+f(x)$ with $k_y$ a constant real number. Notice that this is closely related to a plane wave solution to the field equations, and the possible values of $k_y$ are (dynamically)  constrained by Eq.~\eqref{tosolveBH}. By using $\lambda$ as a parameter, the trajectories $\x=\x(\lambda)$ of light rays are found through the assignment \cite{wald2022}
\begin{equation}
    \k=\frac{\d\x}{\d \lambda}.
\end{equation}
Thus, $k_y=\partial F/\partial y$ implies that $y(\lambda)=y_0+k_y\lambda$, whereas Eq.~\eqref{tosolveBH} gives rise to
\begin{equation}
    \dot{x}=f^{(\pm)}=\alpha\omega\pm\sqrt{n^2\omega^2-k_y^2},\label{eqdifBH}
\end{equation}
with $\dot{x}=\d x/\d \lambda$. The admissible values of $k_y$ are such that $\dot{x}$ remains a real function, that is, $k_y^2 < n^2 \omega^2$. Two possible light rays will exist locally whenever this condition is satisfied.

Notice that far from the slab, $|x|\rightarrow \infty$, where $\alpha$ vanishes, Eq.~\eqref{eqdifBH} implies that $\dot{x}$ assumes two possible values -- one positive and one negative -- corresponding to rightward and leftward ray propagation, respectively. Solving for $\dot{x}=0$ then reveals that
\begin{equation}
    \omega^2(n^2-\alpha^2)<k_y^2<n^2\omega^2\label{constraintBH}
\end{equation}
is the mathematical constraint for a material region to admit one way propagation, i.e., the two solutions for $\dot{x}$ are real and have the same sign. Equation \eqref{constraintBH} has an interesting consequence. For any value of $\alpha\neq0$ in the slab it is always possible to adjust $k_y$ or, equivalently, the angle of incidence of the light rays on the slab, such that the effect occurs.

A sufficiently interesting family of analogs can be found by using the profiles
\begin{align}
    \alpha(x)&=\alpha_0 e^{-x^2/\ell^2},\label{alpha}\\
    n(x)&=n_0\left(1+\Delta  e^{-x^2/\ell^2}\right)^{1/2}.\label{nprofile}
\end{align}
Here $\ell$ models the width of the slab. For this model the slab is concentrated around the plane $x=0$, with $|x|\gg\ell$ modeling light ray propagation in the homogeneous ordinary material with refractive index $n_0$.

\subsection{An analog with horizon-like structures}

\begin{figure}[h!]
\center
 \includegraphics[width=0.45\textwidth]{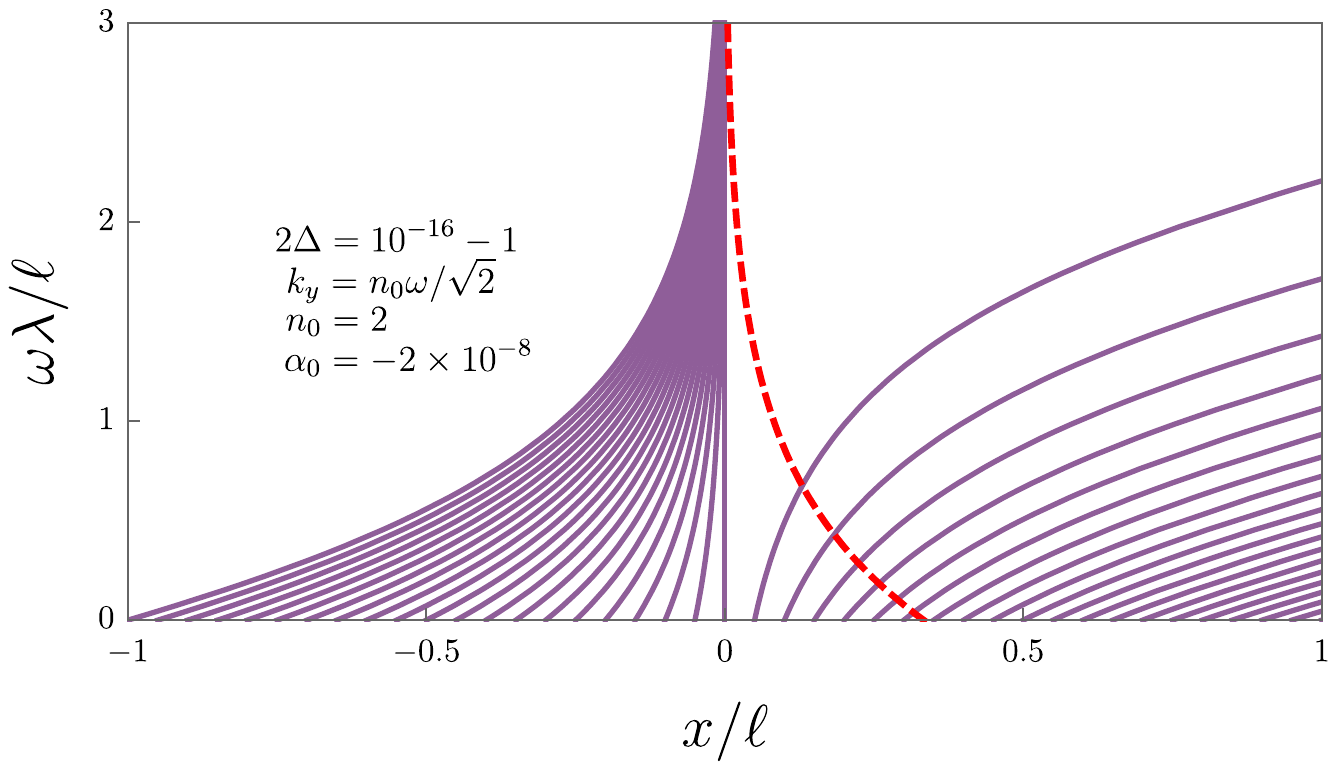}
\includegraphics[width=0.45\textwidth]{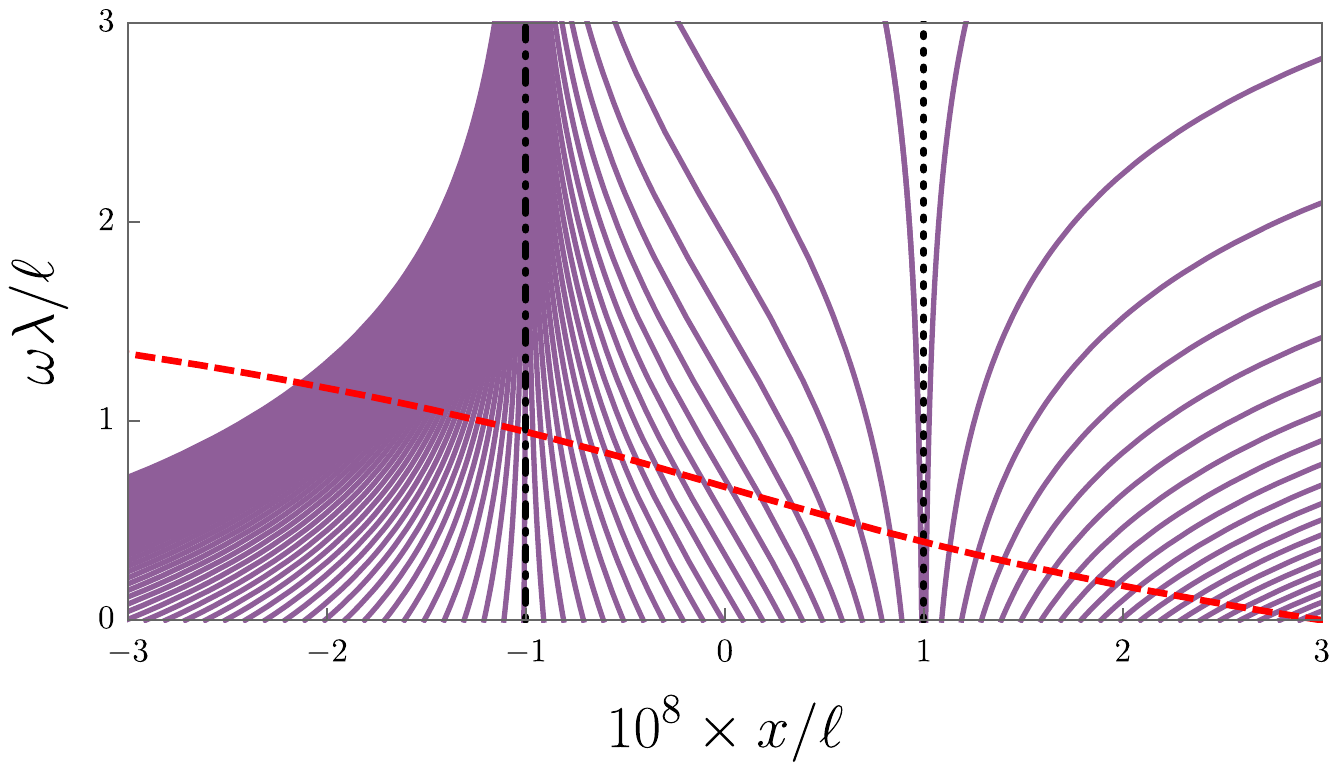}
\caption{Light ray propagation in an analog model displaying event horizons. Top: The dashed red curve corresponds to a single light ray obtained from the negative branch, $f^{(-)}$, in Eq.~\eqref{eqdifBH}. All such rays have the same property: they propagate leftwards and cross the region $x=0$ with a small phase velocity. The rays determined by the positive branch $f^{(+)}$ possess a more interesting behavior. Rays determined by this branch cannot cross the surface $x=0$. Bottom: Geodesics close to $x=0$. Two distinct surfaces at $x\approx\pm 10^{-8} \ell$, determined by $\dot{x}=0$, separate the space into three regions with distinct properties. Light rays in $x<-10^{-8}\ell$ propagate rightwards and accumulate at $x=-10^{-8}\ell$, whereas rays in $10^{-8}\ell<x$ also propagate rightwards but scape to $x\rightarrow\infty$. In the region $|x|<10^{-8}\ell$ light rays propagate only leftwards, and are dragged towards the surface $x=-10^{-8}\ell$. The depicted geodesic profile resembles the principal null geodesics of a Kerr black hole \cite{Carter1968,Gralla}. The surfaces $x=\pm10^{-8}\ell$ play the role of analog event horizons, in the sense that they can only be crossed by light in one direction.
}
\label{fig1}
\end{figure}

In view of Eqs.~\eqref{constraintBH}, \eqref{alpha}, and \eqref{nprofile}, a sufficient condition for the occurrence of a one-way propagation surface is that $k_y$ satisfies
\begin{align}
    1+\Delta-\frac{\alpha_0^2}{n_0^2}<\frac{k_y^2}{n_0^2\omega^2}<1+\Delta,
\end{align}
which, in principle, is always attainable if $k_y^2/(n_0^2\omega^2)$ is sufficiently close to $1+\Delta$. In Fig.~\ref{fig1} we depict a family of geodesics for an effective spacetime, for which light rays can only cross the slab in one direction. The parameters are such that $\alpha_0=- 10^{-8}n_0$, which might be achievable with currently available materials \cite{Zhai2017,Oh2014,Fujii2022}. For the parameters in the figure, $k_y$ is taken to be $n_0\omega/\sqrt{2}$. Therefore, the eikonal equation \eqref{tosolveBH} implies that, far from the slab, light rays with such $k_y$ have $k_x=\pm n_0\omega/\sqrt{2}$, and thus they reach the slab at an angle of $\pi/4$. Naturally, different incidence angles can be found for different material configurations.

Let us summarize the most interesting features depicted in Fig.~\ref{fig1}. Light rays propagating according to the branch $f^{(-)}$ of Eq.~\eqref{eqdifBH} always propagate leftwards and cross the slab at $x=0$ with a smaller phase velocity, i.e., $|k_x|$ has a global minimum (for this configuration) at $x=0$.   
In fact, in the scale of Fig.~\ref{fig1} top panel, such a ray is represented by the dashed curve, that slows down as it approaches $x=0$. The bottom panel depicts a geodesic (dashed curve) crossing the analog horizons. 

The light rays determined by the branch $f^{(+)}$, the continuous lines in Fig.~\ref{fig1}, display a more intricate behavior. Note that these rays cannot cross the two surfaces $x\approx\pm10^{-8}\ell$ determined by $f^{(+)}=0$. Therefore, these surfaces play the role of analog event horizons, to the extent that they can only be crossed by light rays propagating in one direction, given by the branch $f^{(-)}$. Furthermore, all the light rays determined by $f^{(+)}$ in the region $x<10^{-8}\ell$ are dragged toward the analog event horizon at $x=-10^{-8}\ell$, whereas the light rays at $x>10^{-8}\ell$ escape to infinity. This (qualitative) behavior is similar to the one observed for the null principal geodesics of a Kerr black hole \cite{Carter1968,Gralla}.

\subsection{Analog spacetime with a time machine}

Let us now explore another example of possible analog effects stemming from the effective metric \eqref{metric_dow}. Consider again a stationary magnetoelectric material at rest in the laboratory frame, for which $\alpha^{\mu}=(0,\boldsymbol{\alpha})$. In coordinates $x^{\mu}=(t,\x)$, the line element $\d s^2_{\rm eff}$ associated with the effective metric \eqref{metric_dow} reads
\begin{equation}
    \frac{\d s^2_{\rm eff}}{\sqrt{n}}=-\frac{1}{n^2}\left(\d t+\boldsymbol{\alpha}\cdot\d\x\right)^2+\d\x^2.
\end{equation}
Notice that this can be put in a diagonal form by using the coordinate $\tau$ defined locally by $\d\tau=\d t+\boldsymbol{\alpha}\cdot\d\x$, and for constant $n$, the analog spacetime is locally flat. This case includes interesting spacetimes linked to topological defects. For instance, in cylindrical coordinates $\{\rho,\theta,z\}$, if a material is constructed such that $n=1$ and $\boldsymbol{\alpha}=(4J/\rho)\hat{\boldsymbol{\theta}}$, then
\begin{equation}
    \d s^2_{\rm eff}=-(\d t+ 4J\d\theta)^2+\d\rho^2+\rho^2\d\theta^2+\d z^2,\label{cosmic}
\end{equation}
which is the line element of a spinning cosmic string spacetime \cite{Kibble:1976sj, DeLorenci:1999sr}. This is an example of a spacetime with exotic properties, like closed time-like curves. Indeed, if $\d t=\d\rho=\d z=0$, then $\d s^2_{\rm eff}=-(16 J^2-\rho^2)\d\theta^2$, and thus within the region characterized by $0<\rho<|4J|$ the vector field $(\partial_\theta)^a$ is time-like. In what follows, we consider an example of a slab-like configuration that can capture similar features.

For the sake of illustration, let us consider a family of analogs in an exotic metamaterial for which the magnetoelectric effects are not necessarily weak or the refractive index can approach zero. We consider in detail light ray propagation perpendicular to the slab plane and for a constant refractive index, i.e., we take $k_y\equiv0$ and $\Delta=0$ in Eq.~\eqref{nprofile}. In this case, the eikonal assumes the form $S=-\omega t+f(x)$, and Eq.~\eqref{tosolveBH} reduces to
\begin{equation}
    k_x=\frac{\d f}{\d x}=\omega(\alpha\pm n).\label{auxbh}
\end{equation}
The condition for a one-way propagation surface to exist, Eq.~\eqref{constraintBH}, then assumes the form $n_0^2-\alpha_0^2<0$, i.e., a material with either strong magnetoelectric effects or near zero refractive index can, in principle, present a horizon-like structure.

Equation \eqref{auxbh} can be integrated for the profile  \eqref{alpha}, producing the exact result for the eikonal
\begin{equation}
    S=-\omega t\pm n_0\omega x+\frac{\sqrt{\pi}\omega\ell}{2}\alpha_0\mbox{erf}\left(\frac{x}{\ell}\right),
\end{equation}
up to an arbitrary integration constant, and where $\mbox{erf}(x/\ell)$ is the error function \cite{Gradius}. In view of Eq.~\eqref{phasecons}, which shows that a geodesic of the curved spacetime is precisely a curve along which the value of $S$ is constant, the geodesics for this model are found by solving $S=S_0$. By using $x$ as a parameter and requiring that $t(x_0)=0$, we find that $t=t^{(\pm)}(x)+t_0$, with
\begin{equation}
    t^{(\pm)}(x)=\pm n_0(x-x_0)+\frac{\sqrt{\pi}\ell}{2}\alpha_0\left[\mbox{erf}\left(\frac{x}{\ell}\right)-\mbox{erf}\left(\frac{x_0}{\ell}\right)\right],\label{lightrayexact}
\end{equation}
are the exact solutions for the light rays in the analog spacetime crossing the event $t=t_0, x=x_0, z=y=0$.
\begin{figure}[h!]
\center
\includegraphics[width=0.45\textwidth]{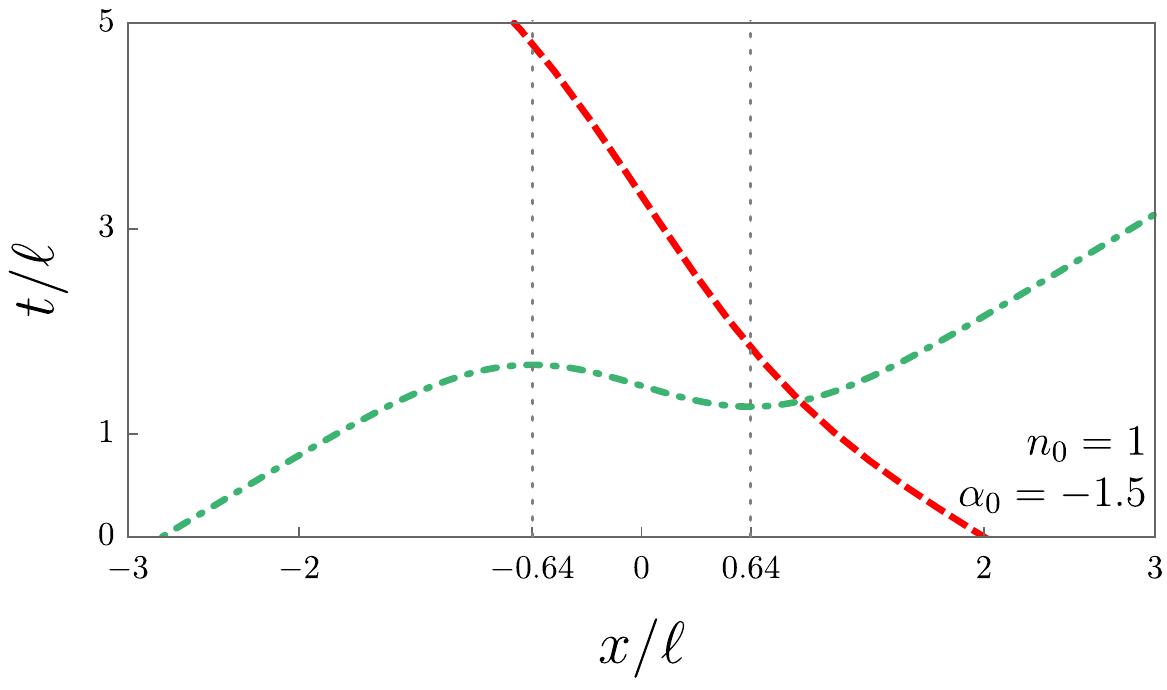}
\caption{Examples of level curves (i.e., light rays) of the eikonal. Here the parameters, $\alpha_0=-1.5, n_0=1$, are such that near $x=0$ the magnitude of the magnetoelectric effects dominate light ray propagation. The green dot-dashed curves represent geodesics determined by the branch $t^{(+)}$, whereas the red dashed curve corresponds to a curve determined by $t^{(-)}$. The phase velocity is given by the inverse of the slopes of the curves, and thus, the light ray modeled by the dashed curve crosses $x=0$ with a smaller phase velocity. For the dot-dashed curve, because the slopes vanish at $x\approx\pm0.64\ell$ (the vertical dotted lines), near these points the phase velocity is superluminal.}
\label{fig2}
\end{figure}
Figure \ref{fig2} depicts several characteristic examples of the light rays given by Eq.~\eqref{lightrayexact} for $n_0=1$ and $\alpha_0=-1.5$. Note that Fig.~\ref{fig2} is a true spacetime diagram depicting the light rays in the hypothetical magnetoelectric material. Accordingly, the inverse of the slopes in Fig.~\ref{fig2} determine the phase velocity of the light rays, which, for the rays modeled by $t^{(-)}$ (dashed curve), diminishes in magnitude near $x=0$.

For the light rays determined by $t^{(+)}$ (dot-dashed curve), because the slopes vanish near the points determined by $\alpha+n=0$, or $x\approx\pm0.64\ell$ for the parameters under study  (vertical dotted curves), we find that near these points the phase velocity diverges. This is consequence of the assumption $|\alpha_0|>|n_0|$ near $x=0$, which might correspond to an inexistent or highly dispersive material, for which the geometric optics approximation adopted here \textit{does not hold}. 

We note an important aspect revealed by Fig.~\ref{fig2}, linked to the correspondence between light ray propagation in the optical medium and in the effective curved space. If a spacetime has a metric given by \eqref{effectivemetric} with the coefficients as in Fig.~\ref{fig2}, then, by definition, the continuous curve of Fig.~\ref{fig2} is the path followed by a light ray, and this ray crosses the region $|x|<0.64\ell$ {\it before} reaching it with respect to the global time coordinate $t$. This analog spacetime shares some similarities with the cosmic string spacetime of Eq.~\eqref{cosmic}. In fact, the effective line element in this case reads $\d s^2_{\rm eff}=-(\d t+\alpha \d x)^2+\d x^2$ ($\d y=\d z=0$), and thus, within the region $|x|<0.64\ell$, the vector field $(\partial_x)^a$ is time-like, just like $(\partial_\theta)^a$ is for the line element of Eq.~\eqref{cosmic}.

The dynamics of the light ray in the optical material, however, is rather different. We recall that the curves in Fig.~\ref{fig2} are simply the curves along which the eikonal remains constant and not the worldline of a light ray. The direction on which the light ray is propagating is really determined by the sign of $\d x/\d t$, where $t$ is the laboratory time. This means that the dot-dashed curve in Fig.~\ref{fig2} corresponds to {\it three} distinct light rays, one propagating rightwards in the region $x<0.64\ell$, a second one propagating leftwards in the region $|x|<0.64\ell$, and a third one propagating rightwards in the region $x>0.64\ell$. This shows that the analogy between light ray propagation in curved spaces and in optical analogs presents subtle nuances that, if not carefully understood, might lead to wrong claims. Also, note that the surface $x=-0.64\ell$ acts as a sink, attracting all nearby light rays, whereas the surface $x=0.64\ell$ repels nearby rays.

\section{Final remarks}
\label{discussion}
There are different ways of producing formal analogies between light propagation in optical media and in curved spacetimes. 
The way explored in this paper is based on the description of light propagation in an optical medium through an effective geometric interpretation, as formally discussed in sections \ref{secmax} and \ref{two}. In such scenario, it is possible to relate the optical coefficients of the medium in consideration with the metric components of a curved space. 
Another way is to start with a metric solution of general relativity and relate the modification of the electromagnetic fields in such curved spacetime with the constitutive relations of a hypothetical optical medium \cite{plebanski,post,gibbons}.
%
%

In summary the major result of our work is the identification of the effective metric \eqref{effectivemetric} from fairly general assumptions for the constitutive relations \eqref{const1}, \eqref{const2}, and the fact that this effective metric determines light ray propagation within the geometric optics approximation. In this sense a couple of remarks regarding the applicability of the findings are in order. First, we note that the constitutive relations assume a non-dispersive regime. Physically this means that memory effects and dissipation, whose manifestation occurs as a dependence of the optical coefficients on the frequency of the light ray, can be neglected. This imposes some experimental nuances, to the extent that the assumed values for the optical coefficients determine the light ray frequency.

A second important remark is linked to the specific analog model built. The effective metric is fairly general, and includes cases, for instance, of flowing magnetoelectric materials, with potential of unveiling remarkable physical phenomena. Nevertheless, the analog model considered assumes a simplified slab configuration at rest, which already presents structures similar to event horizons. We note that this analog can be produced for any real magnetoeletric slab, and apparent experimental limitations, e.g., the light ray's angle of incidence and distance between the event horizons, can be overcome through adjustments of the experimental setup. Ideally, however, metamaterials with stronger magnetoelectric effects would lead to more interesting analogs.

\begin{acknowledgments}
This work was partially supported by Conselho Nacional de Desenvolvimento Cient\'{\i}fico e Tecnol\'ogico under (grant  305272/2019-5), Funda\c{c}\~ao de Amparo \`a Pesquisa do Estado de S\~ao Paulo (grant 23/07013-2), and Funda\c{c}\~ao de Apoio \`a Pesquisa do Distrito
Federal (grant 00193-00002051/2023-14).  
\end{acknowledgments}

\bibliography{refs}

\end{document}